\documentclass[a4paper,11pt]{article}
\usepackage{pos}

\title{The Giant Radio Array for Neutrino Detection}
 \ShortTitle{GRAND}

\author*[a]{Jo\~ao R. T. de Mello Neto }

\affiliation[a]{Universidade Federal do Rio de Janeiro - UFRJ ,\\
  Ilha do Fund\~ao , Rio de Janeiro , Brazil }

\onbehalf{for the GRAND Collaboration} 

\emailAdd{jtmn@if.ufrj.br}

\abstract{Ultra-high-energy cosmic neutrinos (UHE), with energies above 100 PeV, are unparalleled probes of the most energetic astrophysical sources and weak interactions at energies beyond the reach of accelerators. GRAND is an envisioned observatory of UHE particles - neutrinos, cosmic rays, and gamma rays - consisting of 200,000 radio antennas deployed in sub-arrays at different locations worldwide. GRAND aims to detect the radio emission from air showers induced by UHE particle interactions in the atmosphere and underground. For 
neutrinos, it aims to reach a flux sensitivity of $\sim 10^{-10}$ GeV cm$^{-2}$ s$^{-1}$ sr$^{-1}$, with a sub-degree angular resolution, which would allow it to test the smallest predicted diffuse fluxes of UHE neutrinos and to discover point sources. The GRAND Collaboration operates three prototype detector arrays simultaneously: GRAND@Nan\c cay in France, GRANDProto300 in China, and GRAND@Auger in Argentina. The primary purpose of GRAND@Nan\c cay is to serve as a testbench for hardware and triggering systems. On the other hand, GRANDProto300 and GRAND@Auger are exploratory projects that pave the way for future stages of GRAND. GRANDProto300 is being built to demonstrate autonomous radio-detection of inclined air showers and study cosmic rays near the proposed transition between galactic and extragalactic sources. All three arrays are in the commissioning stages.
It is expected that by 2028, the detector units of the final design could be produced and deployed, marking the establishment of two GRAND10k arrays in the Northern and Southern hemispheres.
 We will survey preliminary designs, simulation results, construction plans, and the extensive research program made possible by GRAND.}

\ConferenceLogo{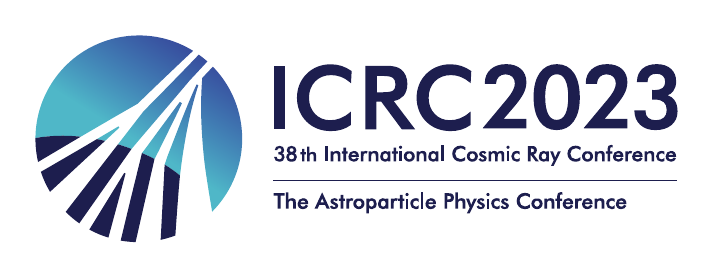}

\FullConference{%
38th International Cosmic Ray Conference (ICRC2023)\\
  26 July - 3 August, 2023\\
  Nagoya, Japan}

\begin{document}
\maketitle

\section{Introduction}   
 
Ultra-high-energy cosmic rays (UHECRs) are atomic nuclei with energies exceeding approximately $10^{18}$ electron volts (eV), and their origins remain a puzzle \cite{AlvesBatista:2019tlv}. The mechanisms by which they attain such extreme energies are still poorly understood. 
Throughout their journey from the point of acceleration to their arrival at Earth, cosmic rays interact with matter and radiation fields along their trajectory, producing numerous secondary particles, including neutrinos and photons. 
This production establishes a significant multi-messenger connection. UHECRs  are deflected from their original paths by intervening magnetic fields and  are attenuated by cosmic photon backgrounds during their propagation.  UHE photons are similarly attenuated.  In contrast, neutrinos travel to Earth largely unaffected by intervening obstacles However, neutrinos, being largely unaffected by obstacles, making them a valuable messenger for exploring the vast reaches   of the high-energy, large-redshift non-thermal Universe.

The Giant Radio Array for Neutrino Detection (GRAND) \cite{GRAND:2018iaj}
 is a proposed large-scale observatory specifically designed to unravel and investigate the sources of UHECRs. One of its primary objectives is to discover and study UHE neutrinos. GRAND will detect the radio signals emitted when UHE cosmic rays, gamma rays, and neutrinos  produce extensive air showers (EAS) in the Earth's atmosphere.
  Its configuration also enables comprehensive fundamental particle physics, cosmology, and radioastronomy studies.
   GRAND will also play a significant role in detecting neutrino emissions  from transient astrophysical sources  \cite{Guepin:2022qpl}.  Below, we will delve into the concept of GRAND, its physics topics, simulated performance,  the ongoing development of prototyping arrays, and the proposed phased implementation. 
 
 \section{The GRAND project}
 \subsection{Radio-detection of UHE neutrinos}
  
 When a cosmic particle interacts with the Earth's atmosphere, it initiates an EAS. This cascade of particles, in turn, produces electromagnetic radiation primarily due to the deflection of charged particles within the shower by the Earth's magnetic field. This phenomenon, known as geomagnetic emission, exhibits coherence in the  tens of MHz frequency range. Consequently, it generates short-duration ($\sim 100$ ns), transient electromagnetic pulses with amplitudes significant enough to enable the detection of the EAS, given that the energy of the shower exceeds approximately $10^{16.5}$ eV \cite{Huege:2016veh, Schroder:2016hrv}. Radio-detection of EAS is a mature technique that benefits from the valuable experience gained through numerous previous experiments, such as AERA, LOFAR, CODALEMA, Tunka-Rex, and TREND, that presented the proof of principle that an antenna array can detect EAS in  stand-alone mode \cite{Charrier:2018fle}. 
 
 Cosmic neutrinos are less likely to be detected through interactions with the atmosphere due to their extremely small interaction cross-section with matter.
 Nevertheless, $\nu_{\tau}$ neutrinos can produce $\tau$ leptons  beneath the Earth's surface via charged-current interactions with rock. Thanks to their considerable range in rock (50 m per PeV of energy before decaying) and short lifetime (0.29 ps), $\tau$ leptons can emerge into the atmosphere and decay, initiating a detectable EAS \cite{Fargion:2000iz}. Only Earth-skimming trajectories allow for such a scenario, since the Earth acts as a barrier to neutrinos with energies surpassing $10^{17}$ eV.
 
 This characteristic proves to be advantageous for radio-detection purposes. Due to relativistic effects, the radio emission becomes highly focused in a forward-directed cone, with its opening defined by the Cherenkov angle $\theta_{\textrm{C}} \le 1^{\circ}$. For vertically incoming showers, this results in a radio footprint on the ground that spans only a few hundred meters in diameter. 
 Consequently, a dense array of antennas is required to sample the signal in this scenario adequately.
  However, for an air shower with highly inclined trajectory, the increased distance between the antennas and the emission zone, coupled with the projection effect of the signal on the ground, leads to a few kilometers-long  footprint \cite{Huege:2016veh, Schroder:2016hrv}. By targeting air showers with such inclined trajectories, it becomes feasible to detect them using a sparser and larger array, typically employing one antenna per square kilometer. This capability serves as a crucial feature of the GRAND detector.
 
 GRAND also incorporates the strategy of selecting mountainous regions with advantageous topographies as deployment sites. An optimal topography involves two parallel mountain ranges spaced several tens of kilometers apart. One range serves as the target for neutrino interactions, while the other functions as a screen onto which the subsequent radio signal is projected. Simulations reveal that such configurations lead to an enhanced detection efficiency, approximately four times greater than that of a flat site  \cite{GRAND:2018iaj}.
 
 \subsection{Simulated performance}
 
 The end-to-end simulation pipeline used to determine the sensitivity of GRAND incorporates the intricate topography of the deployment site and the extensive instrumented area.
   Given the complexity of the task at hand, we ensure the inclusion of all relevant physics while striving to optimize computational performance. We individually validate each simulation stage by comparing it with existing codes. The complete simulation chain is described in detail in \cite{GRAND:2018iaj}.
 
\begin{figure}[ht]
 \centering
 \includegraphics[width=\textwidth]{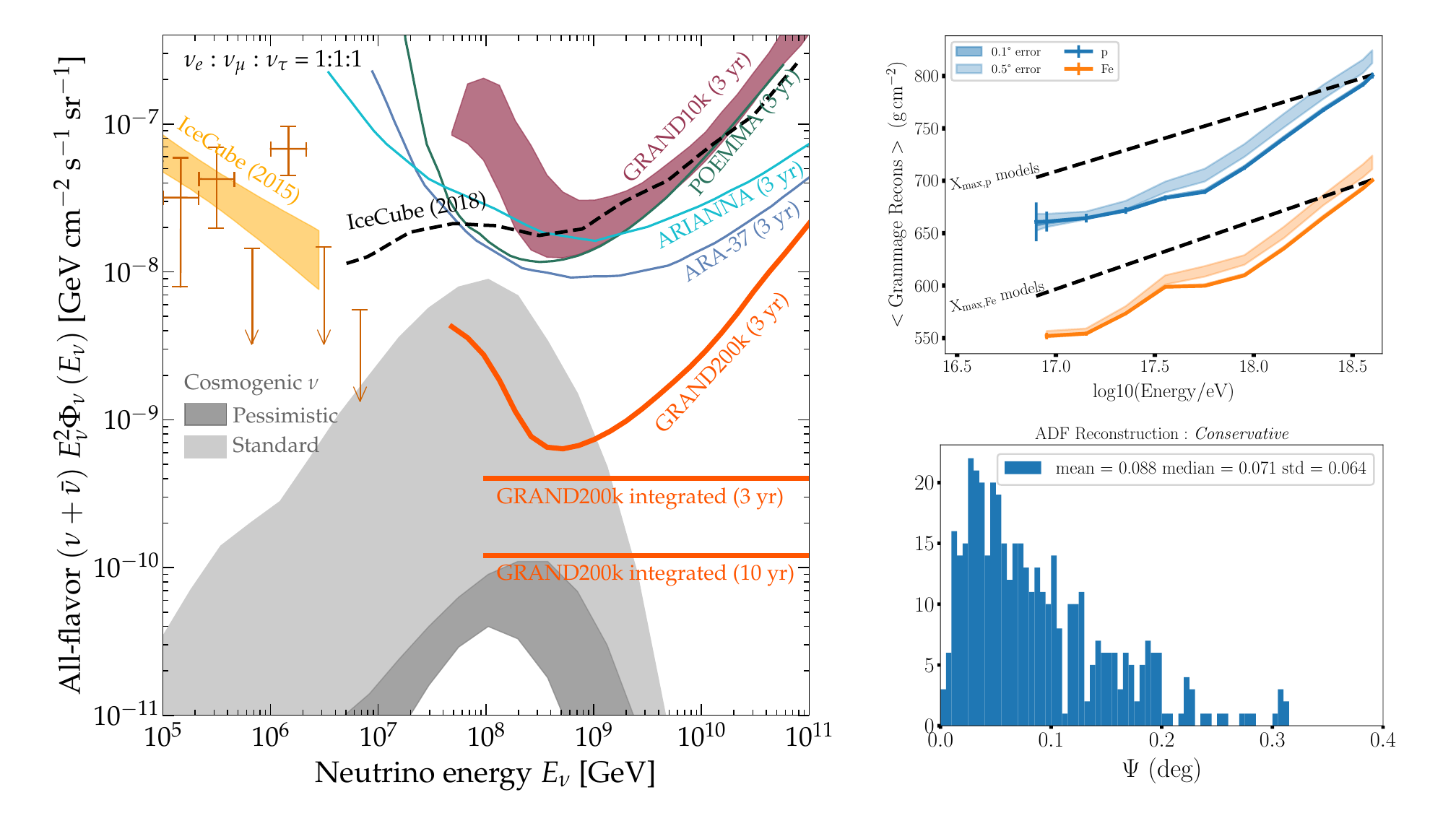}
 \caption{\textit{Left}: Predicted cosmogenic neutrino flux, compared to experimental upper limits and sensitivities. Gray-shaded regions are generated by fitting UHECR simulations to Auger spectral and mass-composition data  \cite{GRAND:2018iaj}. \textit{Upper right}:  Mean value of the radio grammage   ${\{X}_{\textrm e}$ distribution per energy slices for proton  and iron. ${{X}_\textrm{e}}$ can be considered in this model as the (static) point-like source of the radio emission and it is a free parameter of the wavefront model.  The error bars show the statistical fluctuations, taken equal to $\sigma_{{X}_\textrm{e}}/\sqrt{N}$ where  $\sigma_{{X}_\textrm{e}}$ is the standard deviation of the ${X}_\textrm{e}$ values for each simulation and  $N$ is the number of simulated showers per energy slice. The shaded areas correspond to the additional uncertainties associated to error on the direction of origin of the showers for $1-\sigma$ values of $0.1^{\circ}$ and $0.5^{\circ}$, where $\sigma$ is the standard deviation of the  gaussian distribution around the true direction to account for reconstruction errors. \cite{Decoene:2021atj}. \textit{Lower right}: Distributions of the angular distances $\Psi$ for the GRANDProto300-like layout (see section 4) \cite{Decoene:2021ncf}.}  \label{fig1}
 \end{figure}

 The approach described in  \cite{Buitink:2014eqa} has been applied in a preliminary analysis to reconstruct the depth of maximum development of cosmic ray-induced showers  ($X_{\textrm{max}}$)  using  a GRAND-like array. This method achieves $X_{\textrm{max}}$ resolution smaller than 40 g cm$^{-2}$, assuming knowledge of the shower energy and core position \cite{Guepin:2019cmd}. Another study, based on a spherical fit of the wavefront, even though it  does not measure $X_{\textrm{max}}$ directly, uses a figure of merit to estimate that the resolution will be slightly worse than 17~g~cm$^{-2}$ \cite{Decoene:2021atj}. See figure \ref{fig1}, upper right. 
  
Innovative reconstruction techniques  performing fits to the strength of the radio signal as a function of the angle from the shower axis (angular distribution function, ADF) have showcased the potential to achieve angular resolutions of approximately $0.1^{\circ}$ in determining the arrival direction of particles, as shown in figure \ref{fig1}, lower right  \cite{Decoene:2021ncf}.  Even though this result was developed and tested using simulated data only, this level of precision opens up the possibility of conducting neutrino and gamma-ray astronomy with the GRAND observatory.
 
 For neutrino searches, the shower energy of the EAS only provides a lower bound on the initial neutrino energy. The initial findings regarding  energy resolution are promising. Using a preliminary reconstruction 
 based on deep learning methods, an  energy resolution  of 15\% for incoming protons and iron nuclei was achieved without implementing an antenna response and in an idealized scenario for radio-detection \cite{BeatrizThesis}.   
Another preliminary global reconstruction method that uses the angular distribution function also yields  a 20\% energy resolution \cite{Decoene:2021ncf}. Other machine learning and analytical methods are under development within the Collaboration.  Consequently, a final energy resolution of 10\% can likely be attained.

\section{GRAND science case}
\subsection{Ultra-high-energy messengers}
The interaction between UHECRs and the  cosmic microwave background (CMB) and extragalactic background light (EBL) generates {\it cosmogenic} UHE neutrinos (and photons), with energies in excess of $10^{17}$~eV.  Despite our limited understanding of the sources of UHECRs, their existence is assured.
Even with pessimistic assumptions, GRAND has the potential to  discover UHE neutrinos - even if their flux is low. The sensitivity of GRAND to UHE neutrino reaches $ 4\times 10^{-10}$ GeV cm$^{-2}$ s$^{-1}$ sr$^{-1}$  as shown in figure \ref{fig1}, left. 
Cosmogenic neutrino studies indicate that the outcomes of measurements conducted by GRAND will have significant implications for constraining the sources of UHECRs \cite{AlvesBatista:2018zui, Moller:2018isk}. Additionally, they will provide constraints on the proton fraction at ultra-high energies \cite{vanVliet:2019nse}. The remarkable sensitivity of GRAND, coupled with its sub-degree angular resolution, will unlock the potential for conducting UHE neutrino astronomy, enabling the identification of point sources. 

\begin{figure}[ht]
\centering
\includegraphics[width=0.95\textwidth]{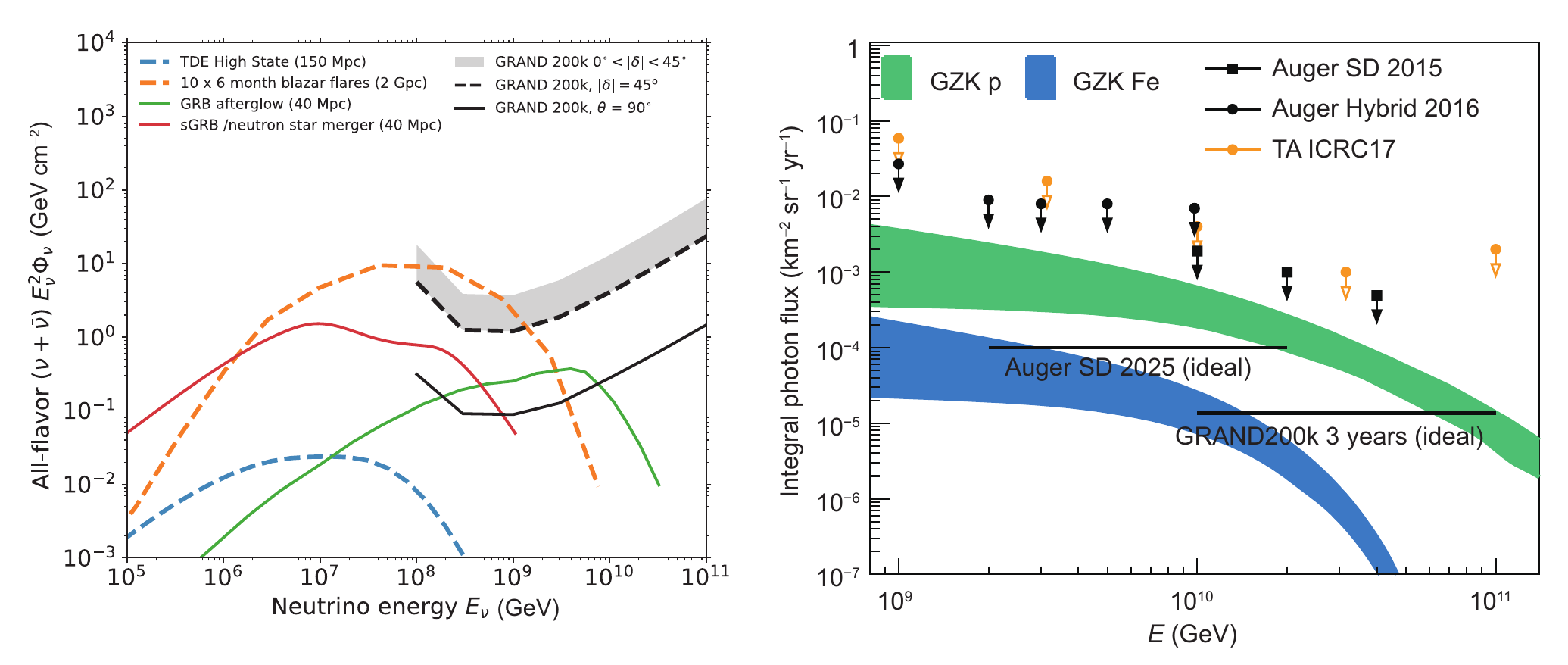}
     \caption{\textit{ Left}:  The sensitivity limits of GRAND for point sources \cite{GRAND:2018iaj}.  It is important to note that these GRAND limits assume the deployment of 200k antennas at a single location. \textit{Right}:  The projected  upper limits of GRAND on the sensitivity to UHE photons after three years of operation are presented. For comparison, we also include the current upper limits from Auger and TA and the projected capabilities of Auger by 2025. Additionally, we overlay the predicted cosmogenic UHE photon flux resulting from pure-proton and pure-iron UHECRs, as estimated in \cite{Denton:2020jft}.}
 \label{fig2}
\end{figure}

Figure \ref{fig2}, left, shows the sensitivity limit of GRAND for point sources. The sources of UHECRs and UHE neutrinos could be distinct.   Therefore, even if a heavy composition is observed in UHECRs, it does not necessarily imply a suppression in the flux of neutrinos at EeV energies \cite{Fang:2016hop}.

Similarly to UHE neutrinos, the cosmogenic flux of UHE photons is also guaranteed. They may be emitted by astrophysical sources, depending on their opacity. However, distant objects cannot be directly observed as they are attenuated  by the CMB and EBL and reprocessed to lower energies.   The most stringent upper limits on UHE photons    can be improved by two orders of magnitude after three years of data-taking by GRAND. See figure \ref{fig2}, right  \cite{GRAND:2018iaj}.

\subsection{Multimessenger astronomy}
With its excellent angular resolution and extensive sky coverage, GRAND has the potential to detect UHE neutrinos associated with transient events in conjunction with electromagnetic emissions.
Using a single array site containing 10,000 antennas (GRAND10k), the instantaneous field of view is about 5\% of the sky; the daily field of view, however, reaches about 80\%.  Using multiple such sites deployed at different locations, as is intended for final configuration of GRAND, the field of view is even larger.
  
GRAND will be a crucial  triggering partner for multimessenger observations, enabling the precise reconstruction of the arrival direction of neutrino-induced air showers near the horizon with sub-degree accuracy and minimal latency. This capability allows GRAND to issue   alerts to other experiments or coordinated systems promptly. Additionally, as a follow-up partner, GRAND can swiftly validate alerts generated by other experiments   as well as gravitational-wave detectors. If the target directions fall within the instantaneous field of view of GRAND, it becomes possible to establish constraints on UHE neutrino emissions originating from a transient.

\section{Experimental setups for prototyping } 
A dedicated design was formulated for the antenna used in the GRAND project, known as the HorizonAntenna  \cite{GRAND:2018iaj}. Its design is optimized to select very inclined EAS.  This antenna features three perpendicular arms, enabling comprehensive polarization measurements of the signal. Positioned at a height of  3.5 m above the ground and optimized for the frequency range of 50--200 MHz, it exhibits optimal sensitivity to near-horizontal signals.  Currently, the GRAND Collaboration has three ongoing prototype arrays taking data. 

\begin{figure}[ht]
\centering
\includegraphics[width=0.95\textwidth]{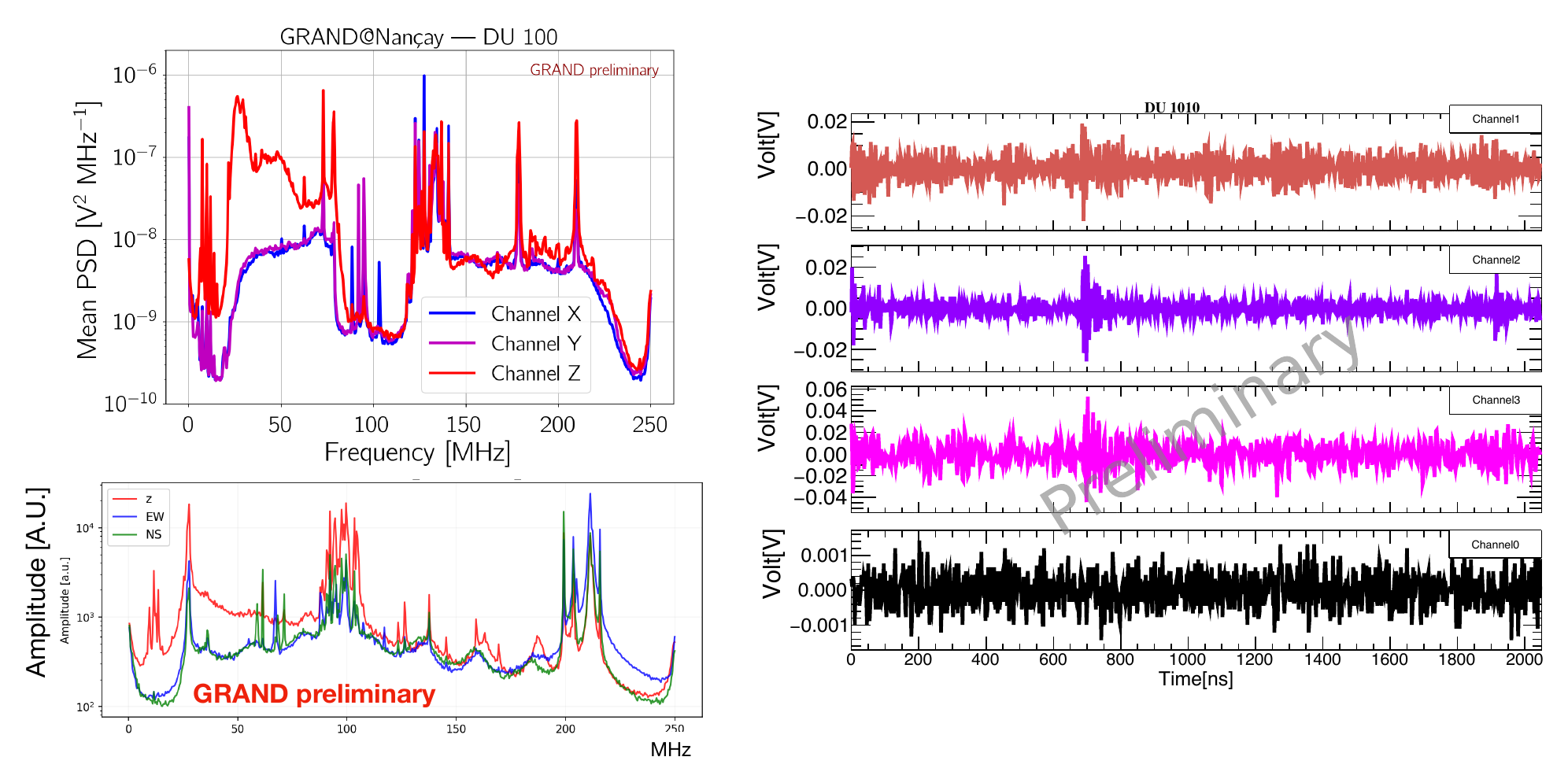}
     \caption{\textit{Upper left}:  Mean power spectrum (PSD) measured at GRAND@Nan\c cay.  A filter is added in the RF acquisition chain of the  setup to cut out FM signals, very strong at that location. \textit{Lower left}:  Average FFT of one detector unit at GRAND@Auger.  \textit{Right}:  Transient event measured in GRANDProto13. Channel 1=X, 2=Y, 3=Z  and channel 0 is free floating. } \label{fig_data}
\end{figure}

Thirteen detection units (GRANDProto13), consisting of antennas and associated electronics, were deployed in the Gobi desert, Gansu Province, China, in February 2023. Data is being collected and analyzed from this initial setup, which serves as the foundation for the GRANDProto300 array \cite{Peng-XiongCRC23}.  One transient event is shown in figure \ref{fig_data}, right.  We are validating the detector unit  design with the ones  already deployed, and once this is done, we will deploy the remaining 70 units already built. This array should already be enough to detect cosmic rays. In the near future, we will build and deploy the remaining  200 units to form GRANDProto300. 

The addition of particle detectors to the prototype array is still under consideration. By utilizing the GRANDProto300 array, we will have the capability to investigate cosmic rays within the energy range of $10^{16.5}$ to $10^{18}$ eV, which encompasses the transitional region between between Galactic and extragalactic UHECRs.  The array will also allow for the detection of radio transients. Moreover, if particle detectors are used, they may help to address the discrepancies observed between simulations and measurements of muons  \cite{muonProb}.

Four detection units were deployed at the Nan\c cay Radio Observatory in France in the autumn of 2022 \cite{PabloICRC23}. The primary objective of the GRAND@Nan\c cay test array is to conduct hardware and  trigger testing. A preliminary spectrum obtained on-site is shown in figure  \ref{fig_data}, upper left. 

Additionally, ten detection units are  being deployed at the Pierre Auger Observatory site from March to August 2023. The main goal is to perform cross-calibration and validation of reconstruction using coincident events with Auger. By taking an average spectrum 
we could verify the presence of FM radio and TV stations in Malargue, as shown in the preliminary spectrum of  figure \ref{fig_data}, lower left. 

The GRAND Collaboration will complete the deployment and continue the operation of prototype arrays: GRAND@Nan\c cay, GRANDProto300, and GRAND@Auger. Meanwhile, it will also focus on characterizing the radio background and exploring the features of autonomous detection of inclined extensive air showers. 

Furthermore the GRAND Collaboration is committed to minimizing its carbon footprint throughout its operations. By implementing sustainable practices and using energy-efficient technologies, we aim to reduce our environmental impact. Efforts are made to optimize the use of resources, reduce waste generation, and promote recycling and reuse. The Collaboration strives to minimize travel-related emissions by employing remote collaboration tools and favoring virtual meetings whenever possible \cite{Aujoux:2021kub}.

\section{Future GRAND timeline } 

We expect that, by 2028, the design of the detector units will be finalized, leading to the construction of two GRAND10k arrays (with 10000 units each). Candidates for the bases of GRAND-North and GRAND-South are China and Argentina, respectively, which will assure a full sky coverage.  Subsequently, in the 2030s, the replication of GRAND10k is expected to commence, resulting in twenty subarrays comprising the entire GRAND project. By scaling up production to an industrial level, the front-end electronics can be transitioned to a fully integrated ASIC design, leading to cost reduction, improved reliability, and greater reproducibility of individual units. Furthermore, the design of each subarray can be customized based on location, topography, or specific scientific objectives.                                                                                                                                                                                                                                                                                     

{ \footnotesize  

}


\begin{thebibliography}{99}
 \setlength{\bibsep}{0.1em}


\bibitem{AlvesBatista:2019tlv}
R.~Alves Batista, J.~Biteau, M.~Bustamante, K.~Dolag, R.~Engel, K.~Fang, K.~H.~Kampert, D.~Kostunin, M.~Mostafa and K.~Murase, \textit{et al.}
``Open Questions in Cosmic-Ray Research at Ultrahigh Energies,''
Front. Astron. Space Sci. \textbf{6}, 23 (2019)

\bibitem{GRAND:2018iaj}
J.~\'Alvarez-Mu\~niz \textit{et al.} [GRAND],
``The Giant Radio Array for Neutrino Detection (GRAND): Science and Design,''
Sci. China Phys. Mech. Astron. \textbf{63}, no.1, 219501 (2020)

\bibitem{Guepin:2022qpl}
C.~Gu\'epin, K.~Kotera and F.~Oikonomou,
``High-energy neutrino transients and the future of multi-messenger astronomy,''
Nature Rev. Phys. \textbf{4}, no.11, 697-712 (2022)

\bibitem{Huege:2016veh}
T.~Huege,
``Radio detection of cosmic ray air showers in the digital era,''
Phys. Rept. \textbf{620}, 1-52 (2016)

\bibitem{Schroder:2016hrv}
F.~G.~Schr\"oder,
``Radio detection of Cosmic-Ray Air Showers and High-Energy Neutrinos,''
Prog. Part. Nucl. Phys. \textbf{93}, 1-68 (2017)

\bibitem{Charrier:2018fle}
D.~Charrier, K.~D.~de Vries, Q.~Gou, J.~Gu, H.~Hu, Y.~Huang, S.~Le Coz, O.~Martineau-Huynh, V.~Niess and T.~Saugrin, \textit{et al.}
Astropart. Phys. \textbf{110}, 15-29 (2019)
doi:10.1016/j.astropartphys.2019.03.002
[arXiv:1810.03070 [astro-ph.HE]].

\bibitem{Fargion:2000iz}
D.~Fargion,
``Discovering Ultra High Energy Neutrinos by Horizontal and Upward tau Air-Showers: Evidences in Terrestrial Gamma Flashes?,''
Astrophys. J. \textbf{570}, 909-925 (2002)

\bibitem{Buitink:2014eqa}
S.~Buitink, A.~Corstanje, J.~E.~Enriquez, H.~Falcke, J.~R.~H\"orandel, T.~Huege, A.~Nelles, J.~P.~Rachen, P.~Schellart and O.~Scholten, \textit{et al.}
``Method for high precision reconstruction of air shower $X_{max}$ using two-dimensional radio intensity profiles,''
Phys. Rev. D \textbf{90}, no.8, 082003 (2014)

\bibitem{Guepin:2019cmd}
C.~Gu\'epin,
``Chasing the cosmic accelerators with high energy astroparticles,''
tel-02411343.


\bibitem{Decoene:2021atj}
V.~Decoene, O.~Martineau-Huynh and M.~Tueros,
``Radio wavefront of very inclined extensive air-showers: A simulation study for extended and sparse radio arrays,''
Astropart. Phys. \textbf{145}, 102779 (2023)


\bibitem{Decoene:2021ncf}
V.~Decoene, O.~Martineau-Huynh, M.~Tueros and S.~Chiche,
``A reconstruction procedure for very inclined extensive air showers based on radio signals,''
PoS \textbf{ICRC2021}, 211 (2021)
doi:10.22323/1.395.0211
[arXiv:2107.03206 [astro-ph.IM]].


\bibitem{BeatrizThesis} B. de Errico, "Deep Learning-based energy reconstruction of cosmic rays with radio emission simulations", Ms. Sc. Thesis, Universidade Federal do Rio de Janeiro (2023), \url{https://pos.if.ufrj.br/wp-content/uploads/2023/07/Masters_dissertation_BeatrizErrico_finalVersion.pdf}


\bibitem{AlvesBatista:2018zui}
R.~Alves Batista, R.~M.~de Almeida, B.~Lago and K.~Kotera,
``Cosmogenic photon and neutrino fluxes in the Auger era,''
JCAP \textbf{01}, 002 (2019)


\bibitem{Moller:2018isk}
K.~M\o{}ller, P.~B.~Denton and I.~Tamborra,
``Cosmogenic Neutrinos Through the GRAND Lens Unveil the Nature of Cosmic Accelerators,''
JCAP \textbf{05}, 047 (2019)

\bibitem{vanVliet:2019nse}
A.~van Vliet, R.~Alves Batista and J.~R.~H\"orandel,
``Determining the fraction of cosmic-ray protons at ultrahigh energies with cosmogenic neutrinos,''
Phys. Rev. D \textbf{100}, no.2, 021302 (2019)

\bibitem{Fang:2016hop}
K.~Fang, K.~Kotera, M.~C.~Miller, K.~Murase and F.~Oikonomou,
``Identifying Ultrahigh-Energy Cosmic-Ray Accelerators with Future Ultrahigh-Energy Neutrino Detectors,''
JCAP \textbf{12}, 017 (2016)

\bibitem{Denton:2020jft}
P.~B.~Denton and Y.~Kini,
``Ultra-High-Energy Tau Neutrino Cross Sections with GRAND and POEMMA,''
Phys. Rev. D \textbf{102}, 123019 (2020)

\bibitem{Peng-XiongCRC23} \textbf{GRAND} Collaboration, P. Ma PoS ICRC2023 (these proceedings) 304


\bibitem{muonProb} L. Cazon \textit{et al.} Working Group Report on the Combined Analysis of Muon Density Measurements from Eight Air Shower Experiments. PoS, ICRC2019:214, (2019)

\bibitem{PabloICRC23} \textbf{GRAND} Collaboration, P. Correa PoS ICRC2023 (these proceedings) 990



\bibitem{Aujoux:2021kub}
C.~Aujoux \textit{et al.} [GRAND],
``Estimating the carbon footprint of the GRAND Project, a multi-decade astrophysics experiment,''
Astropart. Phys. \textbf{131}, 102587 (2021)


\end{thebibliography}
\end{document}